\documentclass{article}

\usepackage[preprint]{tackling_climate_workshop_style}

\usepackage[utf8]{inputenc} 
\usepackage[T1]{fontenc}    
\usepackage{hyperref}       
\usepackage{url}            
\usepackage{booktabs}       
\usepackage{amsfonts}       
\usepackage{nicefrac}       
\usepackage{microtype}      
\usepackage{xcolor}
\usepackage{amsmath}
\usepackage{graphicx}

\title{Contextual Reinforcement Learning for Offshore Wind Farm Bidding}
\author{%
  David L. Cole \\
  Department of Chemical \\
  and Biological Engineering\\
  University of Wisconsin-Madison\\
  Madison, WI 53706 \\
  \texttt{dlcole3@wisc.edu} \\
  \And 
  Himanshu Sharma\\
  Electricity Infrastructure \\
  and Buildings Division\\
  Pacific Northwest National Laboratory\\
  Richland, WA 99354\\
  \texttt{himanshu.sharma@pnnl.gov}\\
  {\bf Corresponding Author}
  \And
  Wei Wang \\
  Electricity Infrastructure \\
  and Buildings Division\\
  Pacific Northwest National Laboratory\\
  Richland, WA 99354\\
  \texttt{w.wang@pnnl.gov}\\
}

\begin{document}

\maketitle

\begin{abstract}
 We propose a framework for applying reinforcement learning to contextual two-stage stochastic optimization and apply this framework to the problem of energy market bidding of an off-shore wind farm. Reinforcement learning could potentially be used to learn close to optimal solutions for first stage variables of a two-stage stochastic program under different contexts. Under the proposed framework, these solutions would be learned without having to solve the full two-stage stochastic program. We present initial results of training using the DDPG algorithm and present intended future steps to improve performance.  
\end{abstract}

\section{Introduction} 
Power systems are a major contributor to climate change because energy production often results in significant greenhouse gas emissions. To combat this challenge, many efforts are being focused on how to integrate, manage, and plan renewable energy (RE) use in the power grid [1, 2, 3]. However, these RE systems (such as wind or solar) can create new difficulties. The energy production of these systems is often stochastic (e.g., there is uncertainty around when or how much the wind will blow). These RE systems also must operate within energy markets which can vary by region and have stochastic prices. Regions typically have two or more markets, often including a long-term--often day-ahead (DA)--market where energy is bid and committed ahead of time, and a short-term, real-time (RT) market, where energy is bid and committed in real time. Markets can compound the challenge of uncertainty, as energy may need to be committed ahead of time without knowing the exact final production of the RE system. Further, power electronic devices (PELs) can greatly influence the operation dynamics in the grid. For example, installing battery storage with a wind farm (WF) can influence how much energy can be committed at different times. To make these RE systems more productive and resilient in the grid, many algorithms, models, and tools are being designed to help make decisions for RE systems in energy markets under different sources of uncertainty [4].

Two-stage stochastic programs (SP) are a common approach for decision making under uncertainty and has been used in many different studies involving RE (see for example, [5, 6, 7]). In two-stage SP, there are a set of first stage (primary) decision variables which must be decided first and a set of second stage (recourse) decision variables that are decided after the realization of the first stage variables. Two-stage SP often considers a large set of scenarios that contain realizations of the uncertain problem data. In this way, the first stage variables are the same across all scenarios, while the second stage variables can vary between scenarios. Scalability is a key challenge of two-stage SP as a large number of scenarios may be necessary for finding meaningful solutions, and the problem size can increase significantly with more scenarios. To combat this challenge, there are a variety of works that seek to simplify or speed up solution of two-stage SP, such as scenario reduction [8, 9], decomposition techniques [10, 11], or through machine learning [12, 13].

In this work, we propose using reinforcement learning (RL) to learn the first stage solution of a contextual 2-stage SP for the day-ahead bidding strategy of an offshore wind farm (WF). Contextual 2-stage SP differs from 2-stage SP in a sense that there is "context" or data that influences the optimal solution but is not a decision variable in the problem (e.g., objective function coefficients). This is motivated in part by the work of Nair et al. [14] who used RL to train a policy to choose first stage solutions of a contextual two-stage SP. However, in their work, all first stage variables were binary while the problem we consider requires continuous first stage variables. Yilmaz and B\"{u}y\"{u}ktahtakin [15] also used RL for 2-stage SP, but they did not incorporate context and used a different setup than we present below. In addition, there are a number of works that use RL for bidding strategies of WFs outside of SP problems [16, 17, 18]. To our knowledge, RL has not been used for learning solutions of contextual 2-stage SP with continuous first stage variables and has not been applied to the offshore WF use case. We present herein our preliminary efforts to build an RL agent and show initial results that suggest that the RL agent starts to learn better actions it should take, but more work is needed to develop a reliable RL agent.

\section{Methods}

The two-stage SP we consider is adapted from [19] where we consider an offshore WF with a battery for storing electricity that can commit to a DA or RT market as shown in \eqref{eq:offshore_wf}.
\begin{subequations}\label{eq:offshore_wf}
    \begin{align}
        \max &\; \sum_{t}^{\mathcal{T}}\lambda^{DA}_t P^{DA}_{t} + \sum_\omega^\Omega p_\omega \sum_t^{\mathcal{T}} \left( \lambda^{RT}_{t, \omega} P^{RT}_{t, \omega} - \lambda^{op}_{t, \omega} P^{op}_{t, \omega} - \lambda^{up}_{t, \omega} P^{up}_{t, \omega}  \right) \label{eq:objective}\\
        \textrm{s.t.} &\; G_{t, \omega} = P^{DA}_t + P^{RT}_{t, \omega} + P^{op}_{t, \omega} - P^{up}_{t, \omega} + P^{ch}_{t, \omega} - P^{dis}_{t, \omega} \quad \forall t \in \mathcal{T}, \omega \in \Omega\label{eq:Pbal}\\
        &\; 0 \le P^{DA}_t, P^{RT}_{t, \omega}, P^{op}_{t, \omega}, P^{up}_{t, \omega}, \quad E^{min} \le E_{t, \omega} \le E^{max} \quad \forall t \in \mathcal{T}, \omega \in \Omega\\
        &\; 0 \le P^{ch}_{t, \omega} \le \bar{P}^{ch}, \quad 0 \le P^{dis}_{t, \omega} \le \bar{P}^{dis} \qquad \forall t \in \mathcal{T}, \omega \in \Omega\\
        &\; E_{t, \omega} - E_{\omega, t - 1} = \eta^{ch} P^{ch}_{t, \omega} - \eta^{dis} P^{dis}_{t, \omega} \qquad \forall t \in \mathcal{T}/\{1, 2\}, \omega \in \Omega\\
        &\; E_{2, \omega} - E_{1} = \eta^{ch} P^{ch}_{t, \omega} - \eta^{dis} P^{dis}_{t, \omega} \qquad \forall \omega \in \Omega\\
        &\; E_{|\mathcal{T}|, \omega} \ge E^{final}, \qquad \forall \omega \in \Omega \label{eq:final_time}
    \end{align}
\end{subequations}
Here, $\mathcal{T}$ is the set of time points in the time horizon (24 hours) and $\Omega$ is the set of scenarios. Decision variables for time $t$ and scenario $\omega$ include the power committed to the DA and RT markets ($P^{DA}_t$ and $P^{RT}_t$), the amounts of under-produced or over-produced power ($P^{up}_{t, \omega}$ and $P^{op}_{t, \omega}$), the power charged or discharged from the battery ($P^{ch}_{t, \omega}$ and $P^{dis}_{t, \omega}$), and the current energy level of the battery ($E_{t, \omega}$). Parameters include the DA and RT market prices ($\lambda^{DA}_t$ and $\lambda^{RT}_t$), the cost to buy power when under-producing ($\lambda^{up}_{t, \omega}$), the cost to curtail power when overproducing ($\lambda^{op}_{t, \omega}$), the total energy produced by the WF in each scenario ($G_{t, \omega}$), maximum charge or discharge levels ($\bar{P}^{ch}$ and $\bar{P}^{dis}$), charge and discharge efficiencies ($\eta^{ch}$ and $\eta^{dis}$), minimum or maximum battery energy levels ($E^{min}$ and $E^{max}$), the initial battery energy level ($E_1$), and a bound on the final battery energy level ($E^{final}$).

\begin{figure}
    \centering
    \includegraphics[width =.8\textwidth]{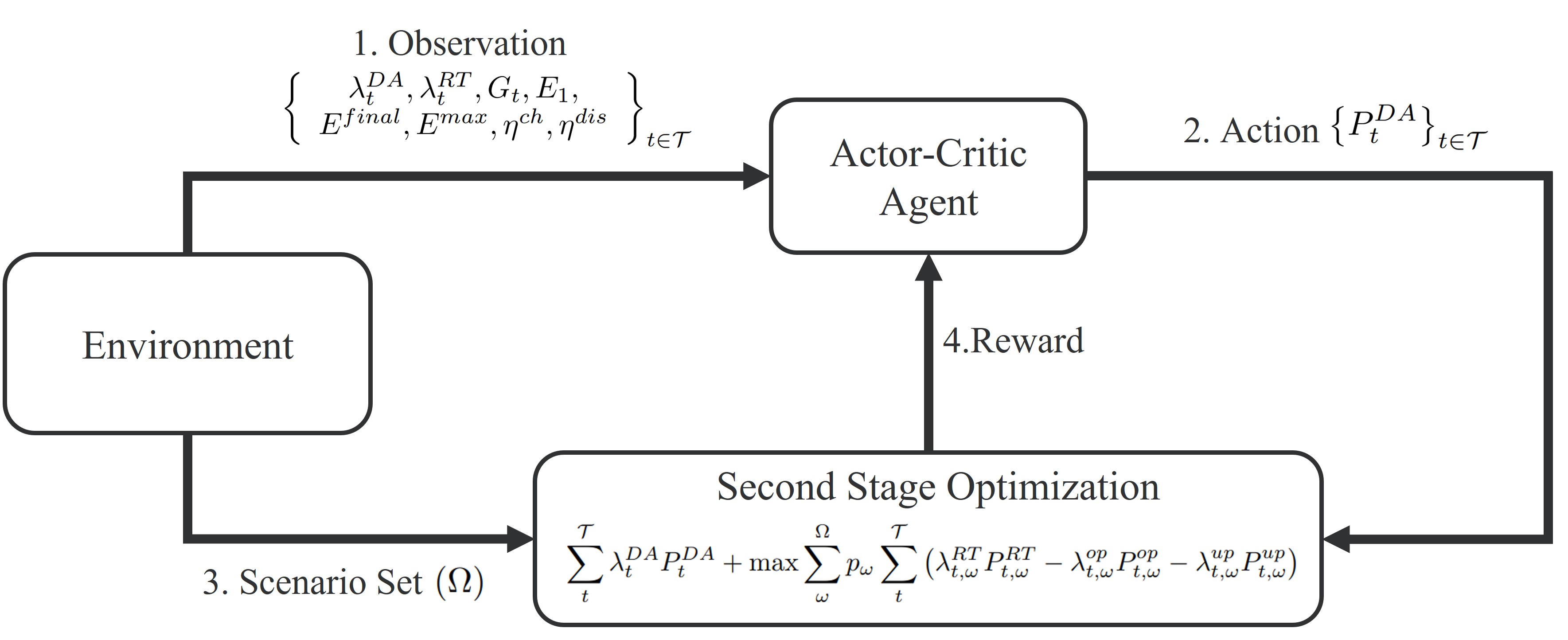}
    \caption{One step of the RL framework for learning the first stage variables. The agent observes the environment, chooses an action, and a reward is computed by solving the second stage of the two-stage SP.}
    \label{fig:RL_framework}
\end{figure}

The RL framework we use is visualized in Figure \ref{fig:RL_framework}. Under this framework, we use RL to predict the first stage variables ($\{P^{DA}_t\}_{t \in \mathcal{T}}$) based on $\{\lambda^{DA}_t\}_{t \in \mathcal{T}}$, $E_1$, forecasts of wind production ($\{G_t\}_{t \in \mathcal{T}}$) and RT prices ($\{\lambda^{RT}_t\}_{t \in \mathcal{T}}$), and based on problem context ($E^{final}$, $E^{max}$, $\eta^{ch}$, and $\eta^{dis}$). We use an actor-critic agent trained using a DDPG algorithm [20] (we also tried the PPO [21] algorithm, but initial performance was better with DDPG). Under this framework, a single step consists of the agent observing the environment ($\{\lambda^{DA}_t, G_t, \lambda^{RT}_t, E_1, E^{final}, E^{max}, \eta^{ch}, \eta^{dis}\}_{t \in \mathcal{T}}$) and predicting an action ($\{P^{DA}_t\}_{t \in \mathcal{T}}$) based on that observation. The action is then passed to the environment, which returns a reward to the agent. Building and training the agent were done using {\tt Stable-baselines3} [22]. While we apply this framework to a specific two-stage SP, the general approach can be extended to other two-stage SPs.

The environment consisted of real price and wind production data, probability distributions relating to the data, and battery characteristics (context). RT and DA price data were for the northeastern United states retrieved from the New England ISO [23], and the wind production data came from [24] (wind and price data came from different locations because we are most interested in a proof of concept for this work). Battery characteristics were randomly generated for each new episode. To compute the reward, the environment solves \eqref{eq:offshore_wf} with $\{P^{DA}_t\}_{t \in \mathcal{T}}$ fixed to be the action of the agent. The optimal value is then divided by $\left( \sum_{\omega \in \Omega} \sum_{t \in \mathcal{T}} G_{t, \omega} \max \left(\lambda^{DA}_t, \lambda^{RT}_{t, \omega} \right) \right) / |\Omega|$ and returned to the agent as the reward (this is effectively a "normalized" reward to account for differences between wind forecasts on different days). For \eqref{eq:offshore_wf}, the environment computes $\Omega$ by sampling from distributions of noise for the RT price data and wind speed data and adding noise to the forecasts to generate unique scenarios (we use $|\Omega| = 10$ and $p_\omega = 0.1, \forall \omega \in \Omega$). To compute the noise distributions, we fit an auto-regressive moving average (ARMA) model to the RT price data and the wind speed data (speeds are converted to power using a power curve), and the residuals (error) in the models were fit to a probability distribution (see [25] and [26]). For RT price data, we used $p = 5$ and $q = 2$, and for wind speed data, we used $p = 3$ and $q = 0$. Lastly, to simplify model training, we set the action space to be $[0, 1]^{|\mathcal{T}|}$. The actions are then multiplied by the forecasted wind power (forcing $0 \le P^{DA}_t \le G_t, \forall t \in \mathcal{T}$). Thus, the agent gives $\{P^{DA}_t\}_{t \in \mathcal{T}}$ by returning a fraction of the wind forecast to commit to the DA market. This theoretically simplifies model training as the model cannot majorly over-commit to the DA market.

\section{Results and Discussion}
We trained the above RL agent for 500,000 time steps and tested the resulting agent. For evaluation, we disconnected the agent from the training loop, generated 2,000 new environments, passed an observation of each environment to the agent, and used the agent's predicted action to compute the expected revenue of \eqref{eq:offshore_wf}. This expected revenue (denoted $f^*_{RL}$) is the optimal solution of \eqref{eq:offshore_wf} with $\{P^{DA}_t\}_{t \in \mathcal{T}}$ fixed to be the action chosen by the agent. For comparison, we also computed the actual solution of $\eqref{eq:offshore_wf}$ without $\{P^{DA}_t\}_{t \in \mathcal{T}}$ being fixed (i.e., the optimal solution of the two-stage SP with the same $\Omega$ as for computing $f^*_{RL}$), and we denote this as $f^*_{SP}$. We also created a simple benchmark bidding strategy where $P^{DA}_t = G_t$ if $\lambda^{DA}_t > \lambda^{RT}_t$ and $P^{DA}_t = 0$ if $\lambda^{DA}_t \le \lambda^{RT}_t, \forall t \in \mathcal{T}$ and solved \eqref{eq:offshore_wf} with $\{P^{DA}_t\}_{t \in \mathcal{T}}$ fixed under this strategy (denoted $f^*_{bench}$). Lastly, we trained an identical RL agent for only 10,000 time steps, used it to predict $\{P^{DA}_t\}_{t \in \mathcal{T}}$, and computed the expected revenue from this agent's decisions to confirm that the original agent is learning (denoted $f^*_{RL'}$).

\begin{figure}
    \centering
    \includegraphics[width = .8\textwidth]{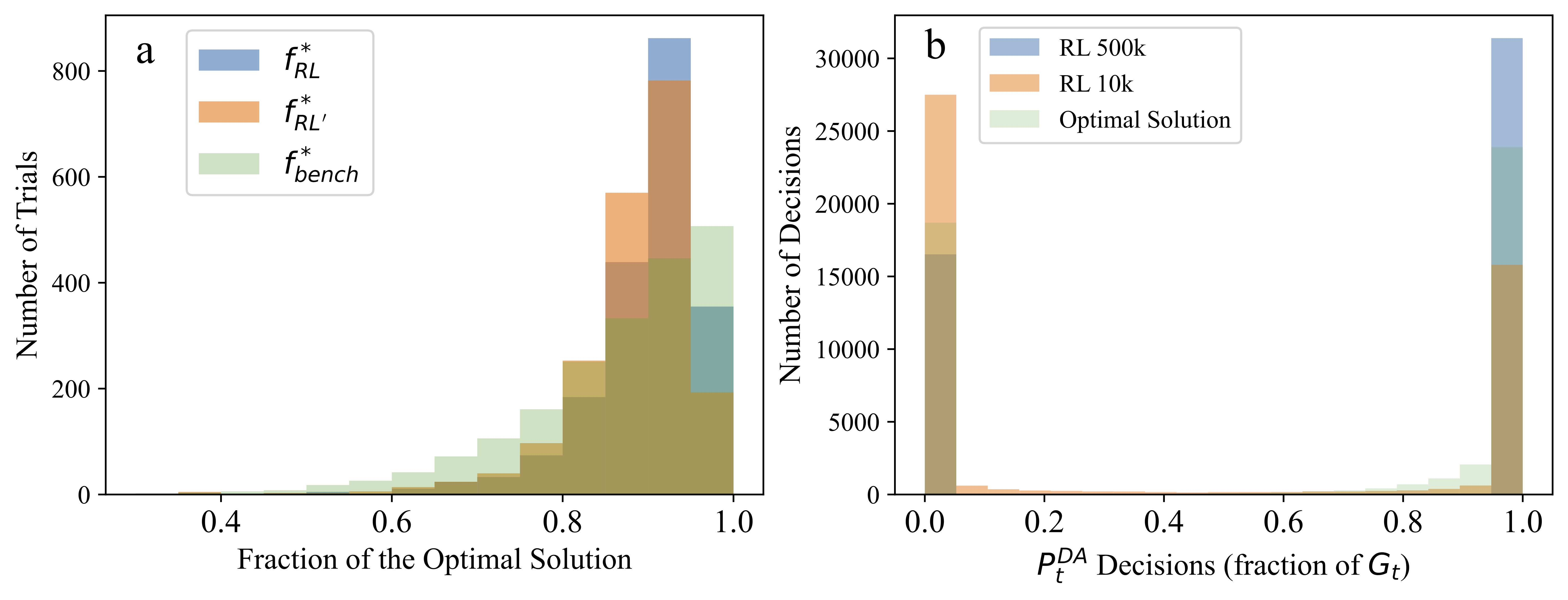}
    \caption{Results from RL agent. (a) shows the distribution of solutions as a fraction of the optimal solution. (b) shows the distribution of decisions chosen by the agent after training for both 10k and 500k steps. Decisions $>1$ for Figure b are considered in the rightmost bar.}
    \label{fig:DDPG_results}
\end{figure}

The results of this test are shown in Figure \ref{fig:DDPG_results}, where Figure \ref{fig:DDPG_results}a shows the distribution of solutions as a fraction of the corresponding $f^*_{SP}$, and Figure \ref{fig:DDPG_results}b shows the distributions of the agent's actions compared with the optimal solutions of the two stage problem. Figure \ref{fig:DDPG_results}a shows that the bidding benchmark problem gets more solutions close (within 95\%) to the optimal solution, but also has more decisions that were $<85\%$ of the optimal solution. On average, the results of the trained RL agent were $\sim1.5\%$ better than the bidding benchmark strategy (or about \$4,000 more per day). Figure \ref{fig:DDPG_results}b suggests that the model is learning the general distribution (the distribution of actions is similar to the optimal values), but {\it when} those actions are taken differs between the RL agent and optimal solution.

Overall, the above RL approach for learning the first stage solutions of a contextual two-stage SP is under-performing what we would hope would be achievable. There are several adjustments that could be pursued in future work which could improve performance. First, we only used 10 scenarios for computing the expected reward, and using more scenarios may be necessary for consistent performance. This could be done with little cost to training time because, once $\{P^{DA}_t\}_{t \in \mathcal{T}}$ are fixed, \eqref{eq:offshore_wf} becomes deterministic and each scenario can be solved independently (i.e., scenarios can be solved in parallel). While we did some initial tuning of hyperparameters using Optuna [27] for the PPO algorithm, further hyperparameter tuning could be explored, particularly for the DDPG algorithm. In addition, changing the architecture of the policy network (we used 3 layers with 16 nodes each) could yield better results as we noticed that the current implementation converges to near-identical actions for many different observations. We are also interested in training for longer than 500,000 time steps (going into the millions of steps) as this may be necessary to learn the entire environment. Further, the reward function may not penalize sub-optimal behavior enough, and our current reward function could be revisited. Lastly, we could also consider the approach of [15], who used two RL agents: one to learn the behavior of the second stage variables and the other to learn the behavior of the first stage variables. This latter approach, while requiring two agents, could be computationally more efficient because no optimization problem needs to be solved.

We recognize that there are both benefits and shortcomings in constructing an RL agent for two-stage SP. First, there is a trade-off between computational cost of training the RL agent compared with the computational cost of solving the two-stage SP. The intention of using RL is to create a model that 1) can be used under a variety of different contexts (which otherwise would each require solving a separate SP), 2) is computationally efficient after training is complete, and 3) does not require ever solving the full two-stage problem. For the second point, this means that the RL agent could be used after training with minimal computational cost and could therefore be used within other large-scale models (e.g., as an input to a larger power grid model or as a surrogate model). For \eqref{eq:offshore_wf}, the RL framework is likely much more expensive than solving the two-stage SP directly because this equation is a linear program which can be efficiently solved, even with a variety of contexts. However, if the above framework is effective, it could likely be applied to more complex problems that are far more difficult to solve (e.g., problems with mixed integer variables) and that have a higher computational cost, making the RL route potentially more desirable. In addition, there is more complexity that could be added to the second stage, such as including other PEL devices (e.g., inverters or converters) which could involve binary variables. In this latter case, the RL framework could be helpful since the second stage problems would be deterministic, meaning several smaller mixed-integer problems can be solved instead of ever solving the larger mixed integer two-stage SP.

In conclusion, we believe that effectively training a RL agent to learn close-to-optimal first-stage variables of a contextual two-stage SP is possible. However further work needs to be done in solving with more scenarios, optimizing hyperparameters and structure, training for more iterations, and revisiting the reward function.

\begin{ack}

This research was supported by the Energy System Co-Design with Multiple Objectives and Power Electronics (E-COMP) Initiative, under the Laboratory  Directed Research and Development (LDRD) Program at Pacific Northwest National Laboratory (PNNL). PNNL is a multi-program national laboratory operated for the U.S. Department of Energy (DOE) by Battelle Memorial Institute under Contract No. DE-AC05-76RL01830. The authors have no competing interests to report.

\end{ack}

\section*{References}

\medskip

\small
[1] Mohlin, K., Bi, A., Brooks, S., Camuzeaux, J., \& Stoek, T. (2019) Turning the corner on US power sector CO2 emissions--a 1990-2015 state level analysis. {\it Environmental Research Letters, 14}(8), 084049.

[2] Osman, A. I., Chen, L., Yang, M., Msigwa, G., Fraghali, M., Fawzy, S., Rooney, D.W., \& Yap, P.S. (2023). Cost, environmental impact, and resilience of renewable energy under a changing climate: a review. {\it Environmental Chemistry Letters, 21}(2), 741-764.

[3] Karunathilake, H., Hewage, K., Mérida, W., \& Sadiq, R. (2019) Renewable energy selection for net-zero energy communities: Life cycle based decision making under uncertainty. {\it Renewable energy, 130,} 558-573.

[4] Zakaria, A., Ismail, F.B., Lipu, M.H., \& Hannan, M.A. (2020) Uncertainty models for stochastic optimization in renewable energy applications. {\it Renewable Energy, 145,} 1543-1571.

[5] Couto, A., Rodrigues, L., Costa, P., Silva, J., \& Estanqueiro, A. (2016) Wind power participation in electricity markets--the role of wind power forecasts. In {\it 2016 IEEE 16th International Conference on Environment and Electrical Engineering (EEEIC)} (pp. 1-6). IEEE. 

[6] Aghajani, A., Kazemzadeh, R., \& Ebrahimi, A. (2018) Optimal energy storage sizing and offering strategy for the presence of wind power plant with energy storage in the electricity market. {\it International Transactions on Electrical Energy Systems,} 28(11), e2621.

[7] Atakan, S., Gangammanavar, H., \& Sen, S. (2022). Towards a sustainable power grid: stochastic hierarchical planning for high renewable integration. {\it European Journal of Operational Research} 302(1), 381-391.

[8] Bertsimas, D., \& Mundru, N. (2023) Optimization-based scenario reduction for data-driven two-stage stochastic optimization. {\it Operations Research, 71}(4), 1343-1361.

[9] Wu, Y., Song, W., Cao, Z., \& Zhang, J. (2021) Learning scenario representation for solving two-stage stochastic integer programs. In {\it International Conference on Learning Representations}.

[10] Soares, J., Canizes, B., Ghazvini, M.A.F., Vale, Z., \& Venayagamoorthy, G.K. (2017) Two-stage stochastic model using Benders' decomposition for large-scale energy resource management in smart grids. {\it IEEE Transactions on Industry Applications, 53}(6), 5905-5914.

[11] Torres, J.J., Li, C., Apap, R.M., \& Grossmann, I.E. (2022) A review on the performance of linear and mixed integer two-stage stochastic programming software. {\it Algorithms, 15}(4), 103.

[12] Bengio, Y., Frejinger, E., Lodi, A., Patel, R., \& Sankaranarayanan, S. (2020) A learning-based algorithm to quickly compute good primal solutions for stochastic integer programs. In {\it Integration of Constraint Programming, Artificial Intelligence, and Operations Research: 17th International Conference, CPAIOR 2020, Vienna, Austria, September 21–24, 2020, Proceedings 17} (pp. 99-111). Springer International Publishing.

[13] Patel, R. M., Dumouchelle, J., Khalil, E., \& Bodur, M. (2022). Neur2SP: Neural two-stage stochastic programming. {\it 36th Conference on Neural Information Processing Systems (NeurIPS 2022)}, 23992-24005.

[14] Nair, V., Dvijotham, D., Dunning, I., \& Vinyals, O. (2018) Learning fast optimizers for contextual stochastic integer programs. In {\it UAI} (pp. 591-600). 

[15] Yilmaz, D. \& B\"{u}y\"{u}ktahtakin, \.{I}.E. (2023). A deep reinforcement learning framework for solving two-stage stochastic programs. {\it Optimization Letters}, 1-28. 

[16] Wei, X., Xiang, Y., Li, J., \& Zhang, X. (2022). Self-dispatch of wind-storage integrated system: a deep reinforcement learning approach. {\it IEEE Transactions on Sustainable Energy, 13}(3), 1861-1864.

[17] Lehna, M., Hoppmann, B., Scholz, C., \& Heinrich, R. (2022) A reinforcement learning approach for the continuous electricity market of Germany: trading from the perspective of a wind park operator. {\it Energy and AI, 8,} 100139

[18] Jeong, J., Kim, S.W., \& Kim, H. (2023) Deep reinforcement learning based real-time renewable energy bidding with battery control. {\it IEEE Transactions on Energy Markets, Policy, and Regulation.}

[19] Kordkheili, R.A., Pourakbari-Kasmaei, M., Lehtonen, M., \& Pouresmaeil, E. (2020) Optimal bidding strategy for offshore wind farms equipped with energy storage in the electricity markets. In {\it IEEE PES Innovative Smart Grid Technologies Europe (ISGT-Europe)} (pp. 859-863). IEEE.

[20] Lillicrap, T.P., Hunt, J.J., Pritzel, A., Heess, N., Erez, T., Tassa, Y., Silver, D., \& Wierstra, D. (2015). Continuous control with deep reinforcement learning. {\it arXiv preprint arXiv:1509.02971}.

[21] Schulman, J., Wolski, F., Dhariwal, P., Radford, A., \& Klimov, O. (2017) Proximal policy optimization algorithms. {\it arXiv preprint arXiv:1707.06347}.

[22] Raffin, A., Hill, A., Gleave, A., Kanervisto, A., Ernestus, M., \& Dormann, N. (2021). Stable-baselines3: reliable reinforcement learning implementations. {\it The Journal of Machine Learning Research, 22}(1), 12348-12355.

[23] ISO New England Inc. (2023). Selectable day-ahead and preliminary real-time hourly LMPs. \url{https://www.iso-ne.com/isoexpress/web/reports/pricing/-/tree/lmp-by-node}. Accessed 9-22-2023.

[24] Grothe, O., K\"{a}chele, F., \& Watermeyer, M. (2022). Analyzing Europe's biggest offshore wind farms: a data set with 40 years of hourly wind speeds and electricity production. \textit{Energies, 15}(5), 1700.

[25] Gangammanavar, H. \& Sen, S. (2016). Two-scale stochastic optimization for controlling distributed storage devices. {\it IEEE Transactions on Smart Grid, 9}(4): 2691-2702. 

[26] Sharma, K.C., Bhakar, R., Tiwari, H.P., \& Chawda, S. (2017). Scenario based uncertainty modeling of electricity market prices. In {\it 2017 6th International Conference on Computer Applications in Electrical Engineering-Recent Advances (CERA)} (pp. 164-168). IEEE.

[27] Akiba, T., Sano, S., Yanase, T., Ohta, T., \& Koyama, M. (2019). Optuna: a next-generation hyperparameter optimization framework. In \textit{Proceedings of the 25th ACM SIGKDD International Conference on Knowledge Discovery \& Data Mining} (pp.2623-2631).

\end{document}